\documentclass[12pt]{iopart}
\usepackage{graphicx}
\usepackage[sorting=none]{biblatex}
\bibliography{article}

\begin{document}

\title[THz transition radiation of electron bunch laser-accelerated in long-scale near-critical density plasmas]{THz transition radiation of electron bunch laser-accelerated in long-scale near-critical density plasmas}

\author{D A Gorlova$^{1,2}$, I N Tsymbalov$^{1,2}$, I P Tsygvintsev$^{3}$, A B  Savelev$^{1,4}$}

\address{$^1$ Faculty of Physics, Lomonosov Moscow State University, 119991, Moscow, Russia}
\address{$^2$Institute for Nuclear Research of Russian Academy of Sciences, 117312, Moscow, Russia}
\address{$^3$Keldysh Institute of Applied Mathematics of Russian Academy of Sciences, 125047 Moscow, Russia}
\address{$^4$Lebedev Physical Institute of Russian Academy of Sciences, 119991, Moscow, Russia}
\ead{gorlova.da14@physics.msu.ru}
\vspace{10pt}
\begin{indented}
\item[]October 2023
\end{indented}

\begin{abstract}
Direct laser electron acceleration in near-critical density plasma produces collimated electron beams with high charge $Q$ (up to $\mu$C). This regime could be of interest for high energy THz radiation generation, as many of the mechanisms have a scaling $\propto Q^2$. In this work we focused specifically on challenges that arise during numerical investigation of transition radiation in such interaction. Detailed analytical calculations that include both diffraction and decoherence effects of characteristics of transition radiation in the THz range were conducted with the input parameters obtained from 3D PIC and hydrodynamic simulations. The calculated characteristics of THz radiation are in good agreement with the experimentally measured ones. Therefore, this approach can be used both to optimize properties of THz radiation and distinguish the transition radiation contribution if several mechanisms of THz radiation generation are considered.
\end{abstract}

\vspace{2pc}

%

\section{Introduction}

Generation of THz radiation from relativistic laser -  plasma interaction is currently being actively investigated \cite{Plasmas2019-vn}. Such THz sources have a unique advantage as they experience no saturation with the increase in laser pulse energy. Therefore, relativistic laser driven THz radiation sources could have pulse energy up to 1\% of main laser pulse energy \cite{Lei2022-ey}, i.e. potentially hundreds of mJ, which cannot be obtained by other generation schemes. This high energy combined with short pulse duration ($\approx$ ps) provides high THz radiation peak field strength, allowing studies of nonlinear THz radiation-matter interactions and several applications in material science \cite{liao2023review}.

A number of THz radiation generation mechanisms in relativistic laser plasma such as transition radiation \cite{Hamster1994-rm}, synchrotron radiation \cite{Liao2020-vh}, sheath radiation \cite{Gopal2012-ry}, linear conversion of plasma wake \cite{Sheng2005-lr}, currents excited via parametric instabilities \cite{Liao2016-yf} had been proposed and investigated. To date, the most widely studied and relevant one being transition radiation (TR), arising when a charged particle crosses electric permittivity discontinuity. In relativistic laser-plasma interaction transition radiation occurs naturally when accelerated electrons cross plasma-vacuum boundary i.e. leave the target. Transition radiation was proposed as the main mechanism for THz generation for various target types: solid targets \cite{Wang2022-jc}, thin foils \cite{liao2019multimillijoule,Liu2019-mc}, gas jets \cite{Van_Tilborg2004-ed,Leemans2003-cq} as well as different micron-sized targets \cite{Dechard2020-pt,Glek2022-vu}. TR was also considered the main mechanism of THz radiation generation in long ($\approx$ hundreds of $\mu$m) optically transparent undercritical ($n_e/n_c\approx 0.1$) plasma in our previous paper \cite{gorlova2022transition}. 

In such long-scale near-critical plasma electrons can be efficiently accelerated with direct laser acceleration (DLA) mechanism both on terawatt \cite{Gahn1999-vv,Tsymbalov2019-kt} and petawatt \cite{Gunther2022-dg} laser setups. Due to high target density, total reported charge of accelerated electrons is up to $\mu C$ \cite{Rosmej2019-dw}, which can generate substantial amount of TR, as this mechanism scales as $\propto Q^2$, where $Q$ - total charge of accelerated electrons. 

In this work we discuss specific challenges that arise with numerical investigation of transition radiation generated by DLA-accelerated electrons in near-critical plasma. Namely, full analytical model, including both diffraction and decoherence effects, is needed to correctly estimate TR characteristics. Multi-stage numerical simulations, consisting of hydrodynamic, PIC and analytical transition radiation calculation, were carried out. The resulting TR properties correspond to experimentally observed ones with high degree of accuracy, which indicates the possibility to estimate numerically the parameters of THz radiation generated by the TR mechanism.

\section{Transition radiation in DLA conditions}

\begin{figure}[ht]
\centering
\includegraphics[scale=0.4]{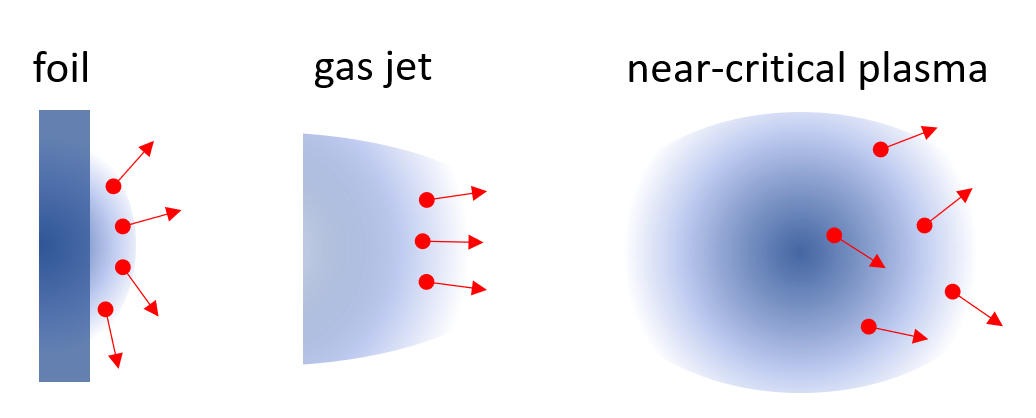}
\caption{Comparison of the formation of transition radiation (TR) for various types of targets used for laser-plasma electron acceleration.}
\label{img:target_comp}
\end{figure}

In the relativistic laser plasma interaction one of the methods for both interpreting experimental data \cite{Liao2016-yf} and predicting the characteristics of a THz radiation source \cite{Zhang2020-kx} is the analysis of THz radiation via particle-in-cell (PIC) simulations. However, due to computational complexity, PIC simulation domain is usually limited to a size of $<$mm$^2$. Therefore, concerning THz radiation, the simulation takes place in the near or in the so-called pre-wave zone \cite{VERZILOV2000135}, where, for example, radiation focusing effects \cite{POTYLITSYN200644} that are not transferred to the far zone can be observed. A strong quasi-static magnetic field of the accelerated electrons, which is difficult to separate from the radiation due to small domain size \cite{Dechard2019-kh}, is also present. These issues are rarely discussed and the analysis of PIC simulation results is usually limited to Fourier filtering of electric field in the low-frequency band \cite{Wang2022-jc,Ding2016-xa}. However, it can lead to a significant overestimation of THz radiation energy, spectral width and field strengths during the PIC simulation analysis. Therefore, it is necessary to develop other approaches to estimate THz radiation parameters. 

As was noted above, generation of TR always occurs during the process of laser-plasma acceleration. The most direct way to evaluate the characteristics of the TR is to calculate them analytically \cite{ginzburg1996radiation}, with the undoubted advantage of pertaining to the far field. Such an analysis had successfully been carried out both for the laser wakefield acceleration (LWFA) \cite{Schroeder2004-zm} and acceleration of electrons in the thin foils \cite{Liao2016-iy}. At the same time, as TR is fully determined by the characteristics of accelerated electrons and the plasma-vacuum boundary, interaction with long-scale near-critical plasma should have unique features. 

The differences in TR generation for different target types are schematically presented in \Fref{img:target_comp}. In the case of electron acceleration in near-critical plasma via DLA electron energy spectrum is exponential with temperature $T=1-15$ MeV. Thus, by the time electrons cross the plasma-vacuum boundary, usually located at a distance of hundreds of microns from the acceleration region, its longitudinal size will be comparable to the THz wavelengths, leading to destructive interference, i.e. temporal decoherence. At the boundary electron beam will also have transverse size on the order of tens of microns (divergence of 0.1-1 rad), leading to spatial decoherence. Thus, both temporal and spatial decoherence will limit TR spectrum at shorter wavelengths. For LWFA decoherence effects are much less pronounced due to the high energy monochromaticity and collimation of the beam \cite{Schroeder2004-zm}. In the case of electron acceleration in thin foils, both the angular and energy distributions of the accelerated electrons can be comparable to the DLA. However, the distance between the electron acceleration region and plasma-vacuum boundary is much smaller and determined by foil thickness (i.e. $\approx$ tens of microns). Thus, temporal decoherence will be suppressed, while spatial decoherence may still be important to consider.

It is also important to take into account diffraction effects due to the limited size of plasma-vacuum boundary. For LWFA diffraction had already been shown to significantly limit THz radiation spectrum at longer wavelengths \cite{Schroeder2004-zm}. For metal foils diffraction are negligible if the rear boundary is unperturbed. For the near-critical plasma, however, diffraction effects must be taken into account as such plasma is usually created through the target heating with an additional prepulse and has a final transverse size of the order of hundreds of microns.

Thus for laser-plasma interaction with long-scale near-critical density plasma no approximations, applicable for other types of targets, can be made. Therefore, it is necessary to take into account effects of both spatial and temporal decoherence, as well as diffraction in analytical calculations. 

\section{PIC simulation of electron acceleration}

Earlier \cite{gorlova2022transition} we had experimentally investigated THz radiation generation in the interaction of 1 TW Ti:Sa laser system (50 mJ, 50 fs, $I=5 \cdot 10^{18}$ W/cm$^2$) with near-critical density preplasma layer with a length of $\approx$ 200 $\mu$m. Here we briefly summarize the results. This layer was created through ablation and subsequent hydrodynamic expansion of 16 $\mu$m mylar tape with an additional Nd:YAG prepulse (200 mJ, 10 ns, $I=10^{12}$ W/cm$^2$). THz radiation parameters were measured for different delays between the main pulse and the prepulse $\Delta t_{fs-ns}$, i.e. different target electron densities, longitudinal and transverse sizes. It was proved that THz radiation is generated via TR mechanism. Properties of TR, calculated with experimentally measured electron beam parameters, corresponded reasonably well to the experimentally measured ones. Note, that in \cite{gorlova2022transition} only diffraction effects were taken into account. Decoherence effects, as we show later, were disguised by the sharp decline of Teflon vacuum-air window transmission for $\nu>3$ THz. 

\begin{figure}[ht]
\centering
\includegraphics[scale=0.5]{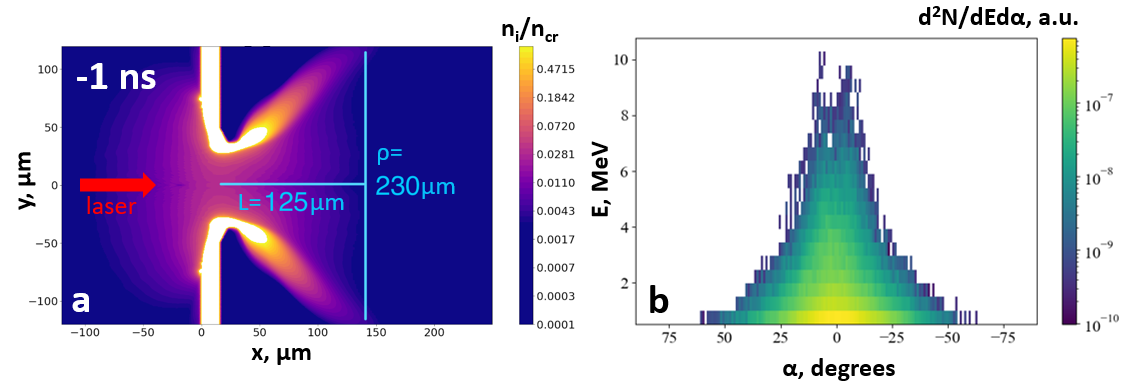}
\caption{Target ion density, obtained in 2D axisymmetric hydrodynamic simulation of interaction of Nd:YAG controlled laser prepulse with 16 $\mu$m thick mylar tape at delay $\Delta t_{fs-ns}=-1$ ns (1 ns after prepulse peak) (a) and energy-angle distribution of accelerated electrons, obtained in 3D PIC simulation for $\Delta t_{fs-ns}=-1$ ns (b). On (a) are also marked $L$ - distance between "point source" of accelerated electrons and plasma-vacuum boundary and $\rho$ - transverse target size, as well as direction of laser propagation. }
\label{img:exp_pic}
\end{figure}

Electron beam parameters that can be measured experimentally (divergence, spectrum, spatial stability) make it possible to take into account only spatial decoherence, while temporal shape of the beam cannot be measured. It can, however, be obtained through particle-in-cell (PIC) simulations. In this work a series of 3D PIC simulations were carried out using the SMILEI \cite{DEROUILLAT2018351} code. Gaussian linearly polarized laser pulse with $a_0=1.5, \tau_{FWHM}=50$ fs was focused to a 4 $\mu$m FWHM spot at a point (x,y)=(10,0) on \Fref{img:exp_pic}a (corresponds to focusing 10 $\mu$m into on the unperturbed target surface). Target profile was obtained through 2D axisymmetric hydrodynamic simulations in the 3DLINE code \cite{krukovskiy20173d}, a more detailed description of this simulation can be found in \cite{ivanov2023laserdriven}. Initially target consisted of neutral carbon atoms with 1 particle per cell (corresponds to 4-6 particle per cell for electrons). PIC simulations grid steps were $\Delta x/\lambda$=1/32, $\Delta y/\lambda, \Delta z/\lambda$=1/4 , $\Delta \tau/t$=1/36. 
Using target profile for $\Delta t_{fs-ns}=-1$ ns (\Fref{img:exp_pic}a), which was established to be optimal for the THz production \cite{gorlova2022transition}, an electron beam with  divergence of $\alpha_{FWHM}=0.12$ rad (\Fref{img:exp_pic}b for E$>$2 MeV) was obtained, which corresponds well to the experimentally observed value of $\approx 0.1$ rad \cite{gorlova2022transition}. Obtained energy-angle distribution of accelerated electrons (\Fref{img:exp_pic}b) was used for analytical calculations of THz radiation. Note that the energy-angle distribution was summed along one of the axes (perpendicular to the polarization direction of the laser pulse) to simplify the calculations.

\section{Framework for analytical TR calculation}

\begin{figure}[ht]
\centering
\includegraphics[scale=0.45]{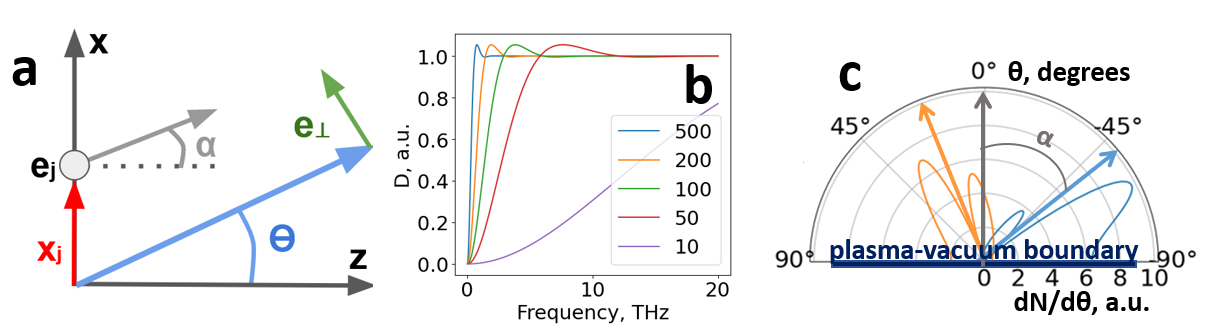}
\caption{(a) - coordinate system used in analytical calculations: z is the axis of electron beam propagation, $\theta$ is the observation angle, $x_j$ is the distance from the $z$ axis at which the $j$th electron crosses the plasma-vacuum boundary at an angle $\alpha_j$, $e_\bot$ - vector perpendicular to the observation direction, (b) - function D (\Eref{eqn:D_func}) that represents the influence of diffraction effects on the THz spectrum, for different transverse sizes of the target (see figure, in $\mu$m) for the $\theta=30^\circ$, electron energy $E=1$ MeV, (c) - angular distribution of the radiation power of one electron with $E=2$ MeV (\Eref{eqn:single_e_dwdwdt}) crossing the plasma-vacuum boundary at an angle of $\alpha_j=20^\circ$ (orange) and $\alpha_j=-50^\circ$ (blue).}
\label{img:calc_parameters}
\end{figure}

As was already mentioned, the most reliable method for calculating TR characteristics is the use of analytical model. Detailed analytical calculations of the TR characteristics had previously been carried out for the LWFA regime \cite{Schroeder2004-zm}. There, the applicability of the sharp plasma-vacuum boundary approximation was demonstrated for gas jet-based laser-plasma accelerators. Here it will also be valid, as the transition region between overcritical (i.e. metal) to undercritical (i.e. vacuum) plasma for the THz radiation occurs on spatial scale $\approx$ several $\mu$m$\ll\lambda_{THz}$ (see \Fref{img:exp_pic}a).

In this work, for the analytical calculation of TR characteristics, a two-dimensional problem was considered; the coordinate system is presented in \Fref{img:calc_parameters}a. In the far field TR, generated at an angle $\theta$ as a result of N electrons crossing the plasma-vacuum boundary can be written as \cite{Schroeder2004-zm}:

\begin{equation}
\label{eqn:e_field_n_electrons}
\vec{E}(\omega,\theta)=i\frac{4\pi e}{\omega}\sum_{j=1}^{N}{\frac{1}{\cos{\theta}}\varepsilon(\theta,u_j,\alpha_j)D(\theta,\omega,u_j,\rho)e^{-ix_jk\sin{\theta}+\phi_j}\vec{e}_\bot} 
\end{equation}
where \cite{Zheng2003-po}:
\begin{equation}
\label{eqn:single_e_dwdwdt}
\varepsilon(\theta,u_j,\alpha_j)=\frac{u_j\cos{\alpha_j}(\sqrt{1+u_j^2}\sin{\theta-u_j\sin{\alpha_j})}}{{(\sqrt{1+u_j^2}-u_j\sin{\theta}\sin{\alpha_j})}^2-{(u_j\cos{\theta}\cos{\alpha_j})}^2}
\end{equation}

is the amplitude of the field of the $j-$th electron with normalized momentum $u_j=\gamma_j\beta_j$, where $\gamma_j=1/\sqrt{1-\beta_j^2}$, $\beta_j=v_j/c$; $\alpha_j$ is the angle and $x_j$ is the transverse coordinate at which the $j-$th electron crosses plasma-vacuum boundary, while $\phi_j$ corresponds to the time delay arising due to electron beam being non-monochromatic. $\varepsilon^2$ corresponds to the known conical distribution of transition radiation with a cone angle $~\frac{1}{\gamma}$ if $\gamma\gg1$.

\begin{equation}
\label{eqn:D_func}
\fl D(\theta,\omega,u_j,\rho)=1-J_0(b_ju_j\sin{\theta)}[b_jK_1(b_j)+\frac{b_j^2}{2}K_0(b_j)]-\frac{b_j^2}{2}K_0(b_j)J_2(b_ju_j\sin{\theta)}
\end{equation}

- function allowing one to take into account the effects of diffraction, arising due to finite transverse size of the plasma-vacuum boundary $\rho$ (see \Fref{img:exp_pic}a), where $b_j=\frac{2\pi\rho}{u_j\lambda}$, $J$ are Bessel functions of the 1st kind, $K$ – Macdonald functions \cite{Schroeder2004-zm}. \Fref{img:calc_parameters}b shows function $D$ for a range of boundary sizes $\rho$. Following notations are also introduced: $\nu$, $\omega$ – ordinary and angular frequencies of radiation, e – electron charge, z – axis of electron propagation, c – speed of light, $ k$ and $\lambda$ are the wave vector and wavelength of radiation with frequency $\omega$, $\vec{e}_\bot$ is the vector, perpendicular to the direction of observation (see \Fref{img:calc_parameters}a). Next, from \Eref{eqn:e_field_n_electrons} we can obtain an expression for the frequency-angular distribution of the energy $W$ of transition radiation:

\begin{equation}
\label{eqn:dW_dwdt}
\frac{d^2W}{d\omega d\theta}=\frac{e^2}{\pi^2c}\sum_{j=1}^{N}\sum_{m=1}^{N}{\varepsilon_j\varepsilon_mD_jD_me^{ik({x_m-x}_j)\sin{\theta} + i(\phi_m-\phi_j)}}
\end{equation}

\Eref{eqn:dW_dwdt} was the main one used for carrying out analytical TR calculations. It includes the effects of diffraction (via the D function, \Eref{eqn:D_func}), electrons crossing the interface at an angle $\alpha$ (via $\varepsilon$, \Eref{eqn:single_e_dwdwdt}) and both spatial (via $x_m-x_j\approx L(\alpha_m-\alpha_j)$ for the small $\alpha$ angles) and temporal (via $\phi_m-\phi_j=\frac{\omega L}{c}(\frac{1}{\beta_m}-\frac{1}{\beta_j})$) decoherence, where $L$ - distance from the acceleration region to plasma-vacuum boundary (see \Fref{img:calc_parameters}a). Further, we did not take into account the coefficient in front of the \Eref{eqn:dW_dwdt}, working in relative units. 

\section{Analytical TR calculations}

The main input parameters of \Eref{eqn:dW_dwdt} are the energy of accelerated electrons E, their divergence $\alpha$, the transverse size of the target $\rho$ and the distance to the interface $L$. The first two parameters were obtained in 3D PIC simulations (\Fref{img:exp_pic}b), while the last two - from hydrodynamic simulations of nanosecond target expansion (\Fref{img:exp_pic}a, $L=$125 $\mu$m, $\rho=$230 $\mu$m). Double summation \Eref{eqn:dW_dwdt} was carried out via simple Python script. 

\begin{figure}[ht]
\centering
\includegraphics[scale=0.4]{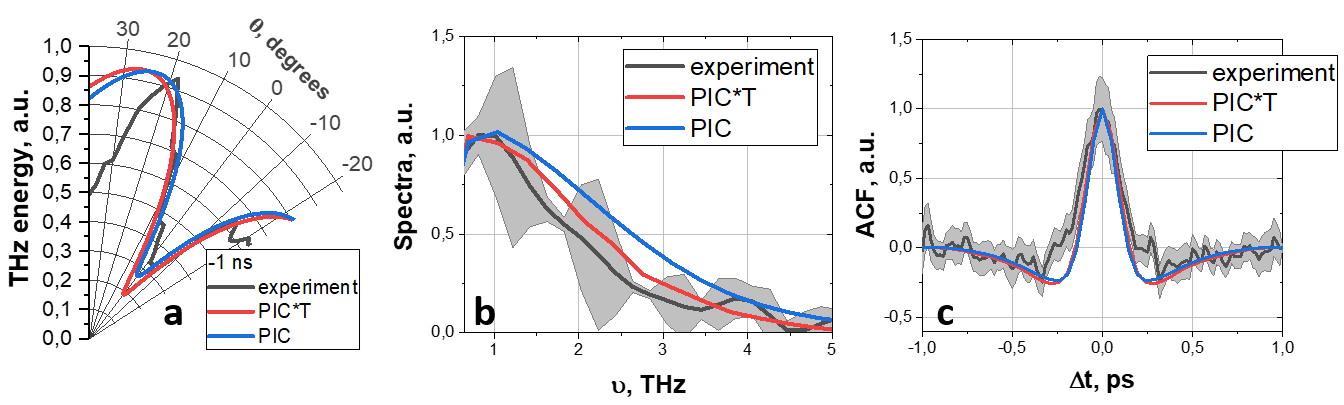}
\caption{Comparison of experimentally measured and numerically calculated THz radiation angular distribution (a), spectrum (b) and autocorrelation function (c) for $\Delta t_{fs-ns}=-1$ ns. Numerically obtained results are shown both with (*T) and without attenuation by Teflon window.}
\label{img:spectra_pic}
\end{figure}

\begin{figure}[ht]
\centering
\includegraphics[scale=0.5]{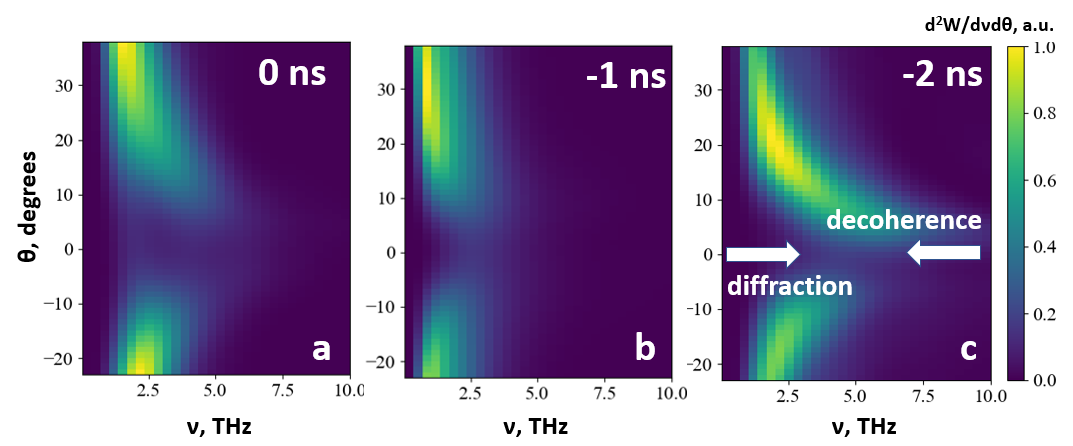}
\caption{Calculated frequency-angular spectra of THz TR for $\Delta t_{fs-ns}$: 0 ns (a), -1 ns (b), -2 ns (c).}
\label{img:spectral_angular_thz}
\end{figure}

The results of such calculation at $\Delta t_{fs-ns}=-1$ ns are presented in \Fref{img:spectra_pic}. It can be seen that numerically calculated angular distribution of transition radiation and its spectrum are in good agreement with experimentally measured ones. Such multi-stage - combination of hydrodynamic, PIC and TR analytical - calculations of TR and their subsequent agreement with experimental results was demonstrated for the first time. Here, calculated THz radiation spectrum and autocorrelation function coincide well with experimentally measured ones (\Fref{img:spectra_pic}b,c). Calculated angular distribution (\Fref{img:spectra_pic}a) is somewhat wider than observed in the experiment which is due to the THz beam being “cut” at the vacuum-to-air Teflon window. As the frequency-angular spectrum of the TR does not change significantly for $\theta>20^\circ$ (see \Fref{img:spectral_angular_thz}b), this had little effect on the experimentally measured spectrum. Note, that accounting for the transmittance of Teflon window has little effect on the observed spectrum (\Fref{img:spectra_pic}b) due to the fact that they have a sharp decline in approximately the same frequency region ($\nu>3$ THz).

\begin{figure}[ht]
\centering
\includegraphics[scale=0.6]{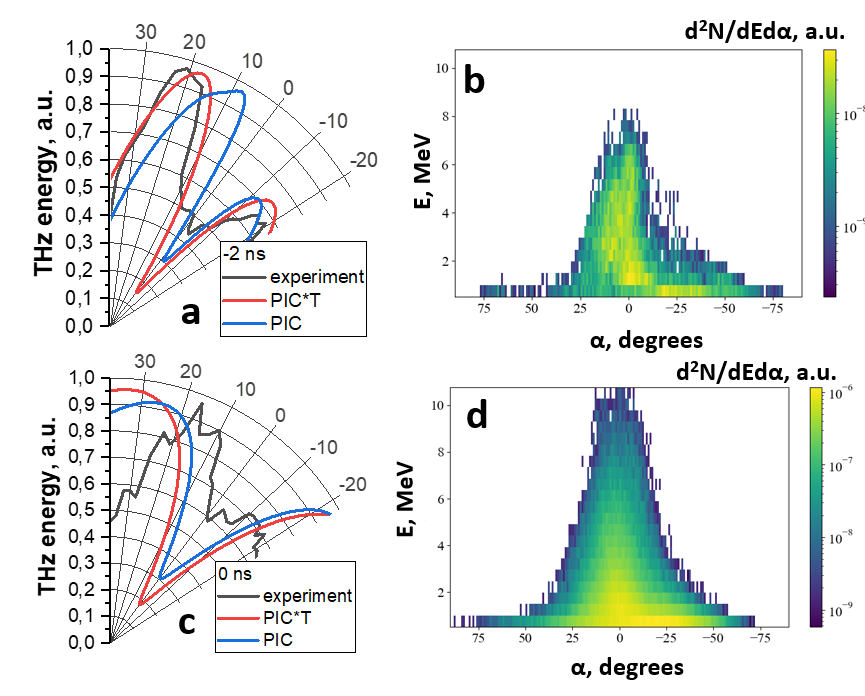}
\caption{Angular distribution of THz radiation measured experimentally and calculated numerically (a,c), as well as energy-angle distribution of accelerated electrons obtained in PIC simulations (b,d), for two values of $\Delta t_{fs-ns}$: - 2 ns (a,b) and 0 ns (c,d). Target parameters for calculations: $\rho$=250 $\mu$m, L=100 $\mu$m ($\Delta t_{fs-ns}=-2$ ns) and $\rho$=150 $\mu$m, L=120 $\mu$m ($\Delta t_{fs-ns}=-2$ ns ). }
\label{img:angular_thz}
\end{figure}

Next, numerical PIC calculations and subsequent calculation of TR characteristics were carried out for neighboring values of $\Delta t_{fs-ns}$, corresponding energy-angle distributions of accelerated electrons and resulting THz angular distributions are presented on \Fref{img:angular_thz}. The angular characteristics of the TR are in agreement with the experimentally measured ones for both $\Delta t_{fs-ns}$. For $\Delta t_{fs-ns}=0$ ns, the experimentally measured angular distribution, similar to \Fref{img:spectra_pic}, experiences a sharp decline, which is not observed in the numerical one is caused by limited acceptance angle of the registration system. However, for $\Delta t_{fs-ns}=-2$ ns, where electrons are accelerated more efficiently and, therefore, generate TR with smaller cone opening angle, a complete coincidence of the angular distributions is observed when Teflon window is taken into account. In addition, the asymmetry of THz radiation angular distribution (\Fref{img:angular_thz}a), which is caused by the asymmetry of the initial distribution of electrons (\Fref{img:angular_thz}b), is also replicated. 

Full (i.e., without taking into account the Teflon window) frequency-angular spectra of TR for all three considered values of $\Delta t_{fs-ns}$ are shown in \Fref{img:spectral_angular_thz}. It can be seen that in the regime of laser plasma interaction under consideration the full spectrum of TR is quite limited: at low frequencies - through diffraction, at high frequencies - through temporal and spatial decoherence. Moreover, varying the parameter $\Delta t_{fs-ns}$ one simultaneously varies electrons energy-angle distribution, the plasma size $\rho$ and the length $L$, leading to non-monotonic changes in TR radiation characteristics. For $\Delta t_{fs-ns}=0$ ns target size is relatively small ($\rho$=150 $\mu$m, L=120 $\mu$m), leading to high diffraction and suppressed decoherence. However, mean electron energies are also small and no effective generation is observed in the frequency range $\nu>5$ THz. Further, for $\Delta t_{fs-ns}=-1$ ns ($\rho$=230 $\mu$m, L=125 $\mu$m) diffraction effects are reduced allowing more efficient TR generation in the frequency range $\nu<2$ THz. For $\Delta t_{fs-ns}=-2$ ns ($\rho$=250 $\mu$m, L=100 $\mu$m) limiting effects of diffraction is similar to $\Delta t_{fs-ns}=-1$ ns. However, due to higher electron energies and lower beam divergence (\Fref{img:angular_thz}b), substantial TR with $\nu>3$ THz is generated.

This interconnection of main parameters defining TR with the change in the main pulse - prepulse delay is a characteristic feature of laser - near-critical density plasma interaction. While it requires careful consideration during analysis of experimental or numerical results, with proper target tailoring it may be possible to use it to control the spectrum of THz radiation. Specifically, a more prominent variation of $L$ is required to shift central TR wavelength efficiently. 

\section{Conclusions}

Generation of transition radiation in the THz frequency range in the interaction of a 1 TW laser pulse with long-scale undercritical plasma layer, created by ablation of a 16 $\mu$m thick mylar tape with a nanosecond pulse, was numerically studied. It was shown that for the parameters of electron beam and target in question is necessary to take into account effects of diffraction and temporal and spatial decoherence during analytical calculation of the TR characteristics. When these effects are taken into account, calculations based on multi-stage numerical simulations (hydrodynamic, PIC and TR analytical) yield THz radiation characteristics (angular distribution, spectrum) that are in good agreement with those measured experimentally. Such complete agreement between numerical calculations and experiments was demonstrated in this work for the first time. It additionally confirms that the mechanism of THz radiation generation in our interaction is transition radiation. It also establishes multi-stage numerical calculations as a viable and sufficiently accurate way to predict the characteristics of THz radiation. It may be of great importance both for establishing mechanisms of THz radiation generation, as well as separating the contribution of TR if various mechanisms are present.

\section{Acknowledgements}
This work was supported by scientific program of the National Center of Physics and Mathematics (project “Physics of high energy density. Stage 2023-2025”). D.G. acknowledges the Foundation for Theoretical Research ‘Basis’ for financial support.
 
\printbibliography

@article{DEROUILLAT2018351,
title = {Smilei : A collaborative, open-source, multi-purpose particle-in-cell code for plasma simulation},
journal = {Computer Physics Communications},
volume = {222},
pages = {351-373},
year = {2018},
issn = {0010-4655},
doi = {https://doi.org/10.1016/j.cpc.2017.09.024},
author = {J. Derouillat and A. Beck and F. Pérez and T. Vinci and M. Chiaramello and A. Grassi and M. Flé and G. Bouchard and I. Plotnikov and N. Aunai and J. Dargent and C. Riconda and M. Grech},
}

@ARTICLE{Plasmas2019-vn,
  title={Review of intense terahertz radiation from relativistic laser-produced plasmas},
  author={Liao, Guo-Qian and Li, Yu-Tong},
  journal={IEEE Transactions on Plasma Science},
  volume={47},
  number={6},
  pages={3002--3008},
  year={2019},
  publisher={IEEE}
}

@ARTICLE{Gopal2012-ry,
  title    = "Observation of energetic terahertz pulses from relativistic solid
              density plasmas",
  author   = "Gopal, A and May, T and Herzer, S and Reinhard, A and Minardi, S
              and Schubert, M and Dillner, U and Pradarutti, B and Polz, J and
              Gaumnitz, T and Kaluza, M C and J{\"a}ckel, O and Riehemann, S
              and Ziegler, W and Gemuend, H P and Meyer, H G and Paulus, G G",
  journal  = "New J. Phys.",
  volume   =  14,
  month    =  aug,
  year     =  2012,
  issn     = "1367-2630",
  doi      = "10.1088/1367-2630/14/8/083012"
}

@ARTICLE{Rosmej2019-dw,
  title     = "Interaction of relativistically intense laser pulses with
               long-scale near critical plasmas for optimization of laser based
               sources of {MeV} electrons and gamma-rays",
  author    = "Rosmej, O N and Andreev, N E and Zaehter, S and Zahn, N and
               Christ, P and Borm, B and Radon, T and Sokolov, A and Pugachev,
               L P and Khaghani, D and Horst, F and Borisenko, N G and
               Sklizkov, G and Pimenov, V G",
  journal   = "New J. Phys.",
  publisher = "IOP Publishing",
  volume    =  21,
  number    =  4,
  pages     = "043044",
  year      =  2019,
  issn      = "1367-2630",
  doi       = "10.1088/1367-2630/ab1047"
}

@ARTICLE{liao2019multimillijoule,
  title={Multimillijoule coherent terahertz bursts from picosecond laser-irradiated metal foils},
  author={Liao, Guoqian and Li, Yutong and Liu, Hao and Scott, Graeme G and Neely, David and Zhang, Yihang and Zhu, Baojun and Zhang, Zhe and Armstrong, Chris and Zemaityte, Egle and others},
  journal={Proceedings of the National Academy of Sciences},
  volume={116},
  number={10},
  pages={3994--3999},
  year={2019},
  publisher={National Acad Sciences}
}

@article{VERZILOV2000135,
title = {Transition radiation in the pre-wave zone},
journal = {Physics Letters A},
volume = {273},
number = {1},
pages = {135-140},
year = {2000},
issn = {0375-9601},
doi = {https://doi.org/10.1016/S0375-9601(00)00486-2},
author = {V.A Verzilov},
}

@misc{ivanov2023laserdriven,
      title={Laser-driven pointed acceleration of electrons with preformed plasma lens}, 
      author={K. Ivanov and D. Gorlova and I. Tsymbalov and I. Tsygvintsev and S. Shulyapov and R. Volkov and A. Savelev},
      year={2023},
      eprint={2309.10530},
      archivePrefix={arXiv},
      primaryClass={physics.plasm-ph}
}

@article{POTYLITSYN200644,
title = {Focusing of transition radiation and diffraction radiation from concave targets},
journal = {Nuclear Instruments and Methods in Physics Research Section B},
volume = {252},
number = {1},
pages = {44-49},
year = {2006},
issn = {0168-583X},
doi = {https://doi.org/10.1016/j.nimb.2006.06.024},
author = {A.P. Potylitsyn and R.O. Rezaev},
}

@article{krukovskiy20173d,
  title={3D simulation of the impact made by a noncentral laser pulse on a spherical tin target},
  author={Krukovskiy, A Yu and Novikov, VG and Tsygvintsev, IP},
  journal={Mathematical Models and Computer Simulations},
  volume={9},
  pages={48--59},
  year={2017},
  publisher={Springer}
}

@article{ginzburg1996radiation,
  title={Radiation by uniformly moving sources (Vavilov--Cherenkov effect, transition radiation, and other phenomena)},
  author={Ginzburg, Vitalii L},
  journal={Physics-Uspekhi},
  volume={39},
  number={10},
  pages={973},
  year={1996},
  publisher={IOP Publishing}
}

@article{liao2023review,
    author = {Liao, Guoqian and Li, Yutong},
    title = "{Perspectives on ultraintense laser-driven terahertz radiation from plasmas}",
    journal = {Physics of Plasmas},
    volume = {30},
    number = {9},
    pages = {090602},
    year = {2023},
    month = {09},
    issn = {1070-664X},
    doi = {10.1063/5.0167730},
}

@article{gorlova2022transition,
  title={Transition radiation in the THz range generated in the relativistic laser—tape target interaction},
  author={Gorlova, D and Tsymbalov, I and Volkov, R and Savel’ev, A},
  journal={Laser Physics Letters},
  volume={19},
  number={7},
  pages={075401},
  year={2022},
  publisher={IOP Publishing}
}

@ARTICLE{Dechard2020-pt,
  title     = "Terahertz emission from submicron solid targets irradiated by
               ultraintense femtosecond laser pulses",
  author    = "D{\'e}chard, J and Davoine, X and Gremillet, L and Berg{\'e}, L",
  journal   = "Phys. Plasmas",
  publisher = "AIP Publishing LLC",
  volume    =  27,
  number    =  9,
  year      =  2020,
  issn      = "1070-664X, 1089-7674",
  doi       = "10.1063/5.0013415"
}

@ARTICLE{Liao2020-vh,
  title     = "Towards {Terawatt-Scale} Spectrally Tunable Terahertz Pulses via
               Relativistic {Laser-Foil} Interactions",
  author    = "Liao, Guo Qian and Liu, Hao and Scott, Graeme G and Zhang, Yi
               Hang and Zhu, Bao Jun and Zhang, Zhe and Li, Yu Tong and
               Armstrong, Chris and Zemaityte, Egle and Bradford, Philip and
               Rusby, Dean R and Neely, David and Huggard, Peter G and Mckenna,
               Paul and Brenner, Ceri M and Woolsey, Nigel C and Wang, Wei Min
               and Sheng, Zheng Ming and Zhang, Jie",
  journal   = "Physical Review X",
  publisher = "American Physical Society",
  volume    =  10,
  number    =  3,
  pages     = "31062",
  year      =  2020,
  issn      = "2160-3308",
  doi       = "10.1103/PhysRevX.10.031062"
}

@ARTICLE{Schroeder2004-zm,
  title    = "Theory of coherent transition radiation generated at a
              plasma-vacuum interface",
  author   = "Schroeder, C B and Esarey, E and van Tilborg, J and Leemans, W P",
  journal  = "Phys. Rev. E",
  volume   =  69,
  number   =  1,
  pages    = "12",
  year     =  2004,
  issn     = "1063-651X",
  doi      = "10.1103/PhysRevE.69.016501"
}

@ARTICLE{Sheng2005-lr,
  title    = "Emission of electromagnetic pulses from laser wakefields through
              linear mode conversion",
  author   = "Sheng, Zheng Ming and Mima, Kunioki and Zhang, Jie and Sanuki,
              Heiji",
  journal  = "Phys. Rev. Lett.",
  volume   =  94,
  number   =  9,
  pages    = "1--4",
  year     =  2005,
  issn     = "0031-9007",
  doi      = "10.1103/PhysRevLett.94.095003"
}

@ARTICLE{Liu2019-mc,
  title    = "Study of backward terahertz radiation from intense picosecond
              laser-solid interactions using a multichannel calorimeter system",
  author   = "Liu, H and Liao, G Q and Zhang, Y H and Zhu, B J and Zhang, Z and
              Li, Y T and Scott, G G and Rusby, D and Armstrong, C and
              Zemaityte, E and Bradford, P and Woolsey, N and Huggard, P and
              McKenna, P and Neely, D",
  journal  = "High Power Laser Science and Engineering",
  volume   =  7,
  pages    = "1--7",
  year     =  2019,
  issn     = "2052-3289",
  doi      = "10.1017/hpl.2018.60"
}

@ARTICLE{Tsymbalov2019-kt,
  title    = "Electrons acceleration in plasma channel in the relativistic
              laser-plasma of solid targets",
  author   = "Tsymbalov, Ivan and Gorlova, Diana and Savel'ev, Andrei",
  journal  = "Proceedings of SPIE - The International Society for Optical
              Engineering",
  volume   =  11037,
  number   = "April 2019",
  pages    = "15",
  year     =  2019,
  issn     = "1996-756X",
  doi      = "10.1117/12.2520767"
}

@ARTICLE{Leemans2003-cq,
  title    = "Observation of Terahertz Emission from a {Laser-Plasma}
              Accelerated Electron Bunch Crossing a {Plasma-Vacuum} Boundary",
  author   = "Leemans, W P and Geddes, C G R and Faure, J and T{\'o}th, Cs and
              van Tilborg, J and Schroeder, C B and Esarey, E and Fubiani, G
              and Auerbach, D and Marcelis, B and Carnahan, M A and Kaindl, R A
              and Byrd, J and Martin, M C",
  journal  = "Phys. Rev. Lett.",
  volume   =  91,
  number   =  7,
  pages    = "1--5",
  year     =  2003,
  issn     = "0031-9007, 1079-7114",
  doi      = "10.1103/PhysRevLett.91.074802"
}

@ARTICLE{Liao2016-yf,
  title    = "Terahertz emission from two-plasmon-decay induced transient
              currents in laser-solid interactions",
  author   = "Liao, G Q and Li, Y T and Li, C and Mondal, S and Hafez, H A and
              Fareed, M A and Ozaki, T and Wang, W M and Sheng, Z M and Zhang,
              J",
  journal  = "Phys. Plasmas",
  volume   =  23,
  number   =  1,
  year     =  2016,
  issn     = "1070-664X, 1089-7674",
  doi      = "10.1063/1.4939605"
}

@ARTICLE{Hamster1994-rm,
  title    = "Short-pulse terahertz radiation from
              high-intensity-laser-produced plasmas",
  author   = "Hamster, H and Sullivan, A and Gordon, S and Falcone, R W",
  journal  = "Physical Review E",
  volume   =  49,
  number   =  1,
  pages    = "671--677",
  year     =  1994,
  issn     = "1063-651X",
  doi      = "10.1103/PhysRevE.49.671"
}

@ARTICLE{Liao2016-iy,
  title    = "Demonstration of Coherent Terahertz Transition Radiation from
              Relativistic {Laser-Solid} Interactions",
  author   = "Liao, Guo Qian and Li, Yu Tong and Zhang, Yi Hang and Liu, Hao
              and Ge, Xu Lei and Yang, Su and Wei, Wen Qing and Yuan, Xiao Hui
              and Deng, Yan Qing and Zhu, Bao Jun and Zhang, Zhe and Wang, Wei
              Min and Sheng, Zheng Ming and Chen, Li Ming and Lu, Xin and Ma,
              Jing Long and Wang, Xuan and Zhang, Jie",
  journal  = "Phys. Rev. Lett.",
  volume   =  116,
  number   =  20,
  pages    = "1--6",
  year     =  2016,
  issn     = "0031-9007, 1079-7114",
  doi      = "10.1103/PhysRevLett.116.205003"
}

@ARTICLE{Ding2016-xa,
  title    = "Sub {GV/cm} terahertz radiation from relativistic laser-solid
              interactions via coherent transition radiation",
  author   = "Ding, W J and Sheng, Z M",
  journal  = "Physical Review E",
  volume   =  93,
  number   =  6,
  year     =  2016,
  issn     = "2470-0053",
  doi      = "10.1103/PhysRevE.93.063204"
}

@ARTICLE{Gahn1999-vv,
  title    = "{Multi-MeV} electron beam generation by direct laser acceleration
              in high-density plasma channels",
  author   = "Gahn, C and Tsakiris, G D and Pukhov, A and Meyer-Ter-Vehn, J and
              Pretzler, G and Thirolf, P and Habs, D and Witte, K J",
  journal  = "Phys. Rev. Lett.",
  volume   =  83,
  number   =  23,
  pages    = "4772--4775",
  year     =  1999,
  issn     = "0031-9007, 1079-7114",
  doi      = "10.1103/PhysRevLett.83.4772"
}

@ARTICLE{Zhang2020-kx,
  title     = "Terahertz radiation enhanced by target ablation during the
               interaction of high intensity laser pulse and micron-thickness
               metal foil",
  author    = "Zhang, Siyuan and Yu, Jinqing and Shou, Yinren and Gong, Zheng
               and Li, Dongyu and Geng, Yixing and Wang, Weimin and Yan,
               Xueqing and Lin, Chen",
  journal   = "Phys. Plasmas",
  publisher = "AIP Publishing LLC",
  volume    =  27,
  number    =  2,
  year      =  2020,
  issn      = "1070-664X, 1089-7674",
  doi       = "10.1063/1.5125611"
}

@ARTICLE{Zheng2003-po,
  title     = "Theoretical study of transition radiation from hot electrons
               generated in the laser--solid interaction",
  author    = "Zheng, Jian and Tanaka, K A and Miyakoshi, T and Kitagawa, Y and
               Kodama, R and Kurahashi, T and Yamanaka, T",
  journal   = "Phys. Plasmas",
  publisher = "American Institute of Physics",
  volume    =  10,
  number    =  7,
  pages     = "2994--3003",
  month     =  jul,
  year      =  2003,
  issn      = "1070-664X",
  doi       = "10.1063/1.1576388"
}

@ARTICLE{Dechard2019-kh,
  title    = "{THz} Generation from Relativistic Plasmas Driven by Near- to
              {Far-Infrared} Laser Pulses",
  author   = "D{\'e}chard, J and Davoine, X and Berg{\'e}, L",
  journal  = "Phys. Rev. Lett.",
  volume   =  123,
  number   =  26,
  pages    = "264801",
  month    =  dec,
  year     =  2019,
  language = "en",
  issn     = "0031-9007, 1079-7114",
  pmid     = "31951438",
  doi      = "10.1103/PhysRevLett.123.264801"
}

@ARTICLE{Wang2022-jc,
  title     = "Correlation of fast electron ejections, terahertz waves, and
               harmonics emitted from plasma mirrors driven by sub-relativistic
               ultrashort laser pulse",
  author    = "Wang, Xiang-Bing and Hu, Guang-Yue and Shen, Bai-Fei and Tang,
               Hui-Bo and Zhang, Zhi-Meng and Gu, Yu-Qiu",
  journal   = "AIP Adv.",
  publisher = "American Institute of Physics",
  volume    =  12,
  number    =  5,
  pages     = "055002",
  month     =  may,
  year      =  2022,
  doi       = "10.1063/5.0077354"
}

@ARTICLE{Glek2022-vu,
  title    = "Enhanced coherent transition radiation from
              midinfrared-laser-driven microplasmas",
  author   = "Glek, P B and Zheltikov, A M",
  journal  = "Sci. Rep.",
  volume   =  12,
  number   =  1,
  pages    = "7660",
  month    =  may,
  year     =  2022,
  language = "en",
  issn     = "2045-2322",
  pmid     = "35538111",
  doi      = "10.1038/s41598-022-10614-0"
}

@ARTICLE{Gunther2022-dg,
  title    = "Forward-looking insights in laser-generated ultra-intense
              $\gamma$-ray and neutron sources for nuclear application and
              science",
  author   = "G{\"u}nther, M M and Rosmej, O N and Tavana, P and Gyrdymov, M
              and Skobliakov, A and Kantsyrev, A and Z{\"a}hter, S and
              Borisenko, N G and Pukhov, A and Andreev, N E",
  journal  = "Nat. Commun.",
  volume   =  13,
  number   =  1,
  pages    = "170",
  month    =  jan,
  year     =  2022,
  language = "en",
  issn     = "2041-1723",
  pmid     = "35013380",
  doi      = "10.1038/s41467-021-27694-7",
  pmc      = "PMC8748949"
}

@ARTICLE{Lei2022-ey,
  title    = "Highly efficient generation of {GV/m-level} terahertz pulses from
              intense femtosecond laser-foil interactions",
  author   = "Lei, Hong-Yi and Sun, Fang-Zheng and Wang, Tian-Ze and Chen, Hao
              and Wang, Dan and Wei, Yan-Yu and Ma, Jing-Long and Liao,
              Guo-Qian and Li, Yu-Tong",
  journal  = "iScience",
  volume   =  25,
  number   =  5,
  pages    = "104336",
  month    =  may,
  year     =  2022,
  language = "en",
  issn     = "2589-0042",
  pmid     = "35602940",
  doi      = "10.1016/j.isci.2022.104336",
  pmc      = "PMC9118729"
}

@ARTICLE{Van_Tilborg2004-ed,
  title     = "Pulse shape and spectrum of coherent diffraction-limited
               transition radiation from electron beams",
  author    = "Van Tilborg, J and Schroeder, C B and Esarey, E and Leemans, W P",
  journal   = "Laser Part. Beams",
  publisher = "Cambridge University Press",
  volume    =  22,
  number    =  4,
  pages     = "415--422",
  month     =  oct,
  year      =  2004,
  issn      = "1469-803X, 0263-0346",
  doi       = "10.1017/S0263034604040078"
}
\end{document}